# Improved critical current densities in $B_4C$ doped $MgB_2$ based wires

## P. Lezza[1,*], C. Senatore[2], R. Flükiger[1,2]

[1] Group of Applied Physics, University of Geneva, 20, rue de l'Ecole de Médecine – CH-1211 Geneva 4, Switzerland

[2] Department of Physics of Condensed Matter (DPMC) and MANEP (NCCR), University of Geneva, 24, quai Ernest Ansermet – CH-1211 Geneva 4, Switzerland

## Abstract

An improvement of the transport critical current density, $J_c$, of $MgB_2$ wires was obtained after addition of 10 wt.% $B_4C$ powders, after reaction at 800°C: $J_c$ values of $1 \cdot 10^4$ A/cm$^2$ at 4.2 K and 9T were obtained for wires of 1.11 mm diameter in a Fe matrix. The starting mixture of Mg and B was doped with sub-micrometric $B_4C$, the ratio being   Mg : B : $B_4C$ = 1 : 2 : 0.08, corresponding to 10 wt.% $B_4C$. For T > 800°C, a decrease of $J_c$ was found, due to the reaction with the Fe sheath. In order to investigate the origin of the improvement of the transport properties for heat treatments up to 800°C, X-ray diffraction measurements were performed. A decrease of the lattice constant $a$ from 3.0854 Å to 3.0797 Å was found, thus suggesting an effect of the substitution of Carbon on the properties of the wires. A comparison with the literature data shows that the addition of $B_4C$ powders leads to the second highest improvement of $J_c$ reported so far after SiC, thus constituting an alternative for future applications.

**Introduction**

The high critical temperature of $MgB_2$ [1] with respect to the industrial low temperature $Nb_3Sn$ and NbTi renders this compound suitable for applications at 20 K. Many groups are currently developing wires and tapes conductors by powder-in-tube procedures, either by the *ex situ* [2,3,4] or by the *in situ* technique [5,6]. The $J_c$ values of properties of *in situ* wires experienced a strong improvement after introducing SiC nanopowders [7], the effect being attributed to the substitution of Boron by Carbon. The effect adding $B_4C$ powders has been studied by various authors [8,9,10,11], who found that B can be partially substituted by C in the $MgB_2$ phase. This was concluded by Yamamoto et al. and by Balaselvi et al. [8,9] from the change of the lattice parameter *a* of $MgB_2$, while Ribeiro et al.[10] reported a decrease of $T_c$. Another author reported higher values of $B_{c2}$ on $MgB_2$ bulk samples after $B_4C$ additions [12]. Since all these works have been performed on bulk samples, we have investigated the possibility of fabricating *in situ* wires by using $B_4C$ submicron powder additions, in order to exploit the potentialities of this compound. We report in the present paper a sizeable increase of $J_c$ after $B_4C$ doping of *in situ* $MgB_2$/Fe wires. The enhancement of $J_c$ is still lower than for SiC additions, but $B_4C$ additions could possibly constitute an alternative in view of an enhancement of the transport properties of $MgB_2$.

**Experimental**

Monofilamentary wires were prepared by the powder-in-tube technique using Mg (99.8% pure, ~325 mesh) and amorphous B (oxygen content: 1.02 %, micrometric particle size) at the stoichiometric ratio, adding 10wt.% of $B_4C$ sub-micron powders (average particle size: 500 nm). The ratio between the starting compounds was chosen to Mg : B : $B_4C$ = 1 : 2 : 0.08, corresponding to 10 wt.% $B_4C$. The precursors were mixed and then inserted in 10 cm long Fe tubes (99.5 % pure) of $\varnothing_{out}$ and $\varnothing_{in}$ of 8 and 5 mm, respectively. The diameter of the tubes was successively reduced applying a combination of swaging and drawing, and wires of 1.11 and 0.657 mm diameter were obtained. The wires were heat treated under Ar flow at various temperatures from 670 to 900°C and for various reaction times. A cross section of a $B_4C$ added $MgB_2$/Fe wire is shown in Fig. 1. For comparison, undoped wires were fabricated with the same procedure up to diameters of 1.11 mm diameter.

The critical current density $J_c$ was measured by the standard 4-point method under applied magnetic field strengths up to 15 T, at 4.2 K and 20 K. X-ray diffraction measurements were performed after extracting the powder from the filaments, using Cu K$\alpha_1$ radiation with a Si standard in order to get a precise indication about the lattice parameter change due to the $B_4C$ addition.

## Results

The XRD patterns of the powder extracted from some of the doped conductors are shown in Fig. 2. At 670°C, the $MgB_2$ superconducting phase has already been formed, as for undoped wires, and is coexisting with boron carbide and some MgO. At 670 and 720 °C, some free Mg and traces of the ternary phase $MgB_2C_2$ are also present, indicating that the reaction is not yet complete.

With higher reaction temperatures, the intensities of the peaks corresponding to $B_4C$, $MgB_2C_2$ and Mg are progressively reduced, leaving only the reflections due to $MgB_2$ and MgO. Some Fe peaks, present in all patterns, are due to residual Fe after extracting the powder from the matrix. These data are compatible with other works [8,9,10,11] which revealed the same secondary phases in the superconducting powder. It has to be noted that the presence of the Fe sheath and the excess of B in the starting mixture have an influence on the kinetics of the reaction [13]. A cross section of the wire after 1 h at 800°C using an optical microscope (Fig. 3) shows a reaction layer of ~8 μm thickness [13,14,15], confirming a partial interaction of the superconducting filament with the sheath. Table 1 shows the variation of the lattice parameter $a$ and $c$ as a function of the reaction temperature. This variation shown in Fig. 4 for the parameter $a$ suggests that the reaction between $B_4C$ with the two main elements leads to a substitution of the Boron by Carbon in the $ab$ planes [8-12,16,17]. Comparing the lattice parameters with those of C doped bulk samples [8,17], it is concluded that the amount of C substituting the B atoms is lower than the nominal one $Mg : B : B_4C = 1 : 2 : 0.08$, but that the substituted C content increases with reaction temperature.

Measurements of the critical current density $J_c$ vs. B were performed at 4.2 K for various reaction conditions. For each set of temperature and reaction time, several wires were measured, in order to verify the homogeneity inside the wires. The results are shown in Fig. 5 where the critical current density $J_c$ of the pure and the $B_4C$ doped wires of 1.11 mm diameter are compared. As expected, the critical current densities of the undoped wire become progressively smaller with higher reaction temperatures. This is in contrast to $B_4C$ doped wires, where an enhancement of $J_c$ with the reaction temperature is observed, in agreement with the results on bulk samples [8]. The highest value measured so far for a wire with 10wt.% $B_4C$ is $1 \cdot 10^4$ A/cm$^2$ at 9 T. The effect of a further enhancement of the reaction temperature on $J_c$ of a $B_4C$ doped wire is shown in Fig. 6. After longer reaction times at 800°C, only a small reduction of the critical current density $J_c$ is observed, in contrast to a considerably larger reduction at 900°C.

In order to confirm these results, additional wires doped with 10wt.% $B_4C$ submicron powder particles were fabricated, again with diameter of 1.11 mm and 0.657 mm. Measurements of the critical current density vs. B were performed both at 4.2 and at 20 K and are shown in Fig.

7. The results confirm the data of the first series of wires, with a $J_c$ value of $10^4$ A/cm$^2$ at 4.2K and 9 T. The same Jc value was obtained at 20K and 4 T.

**Discussion**

The presently fabricated wires show an improvement of the transport $J_c$ of MgB$_2$ wires when adding 10 wt.% B$_4$C submicron powders, at reaction temperature as low as 800°C. A comparison with other $J_c$ values is shown in Fig. 8, showing a sizeable increase of $J_c$ when comparing to the B$_4$C doped bulk samples by Yamamoto [8]. These authors [8] started with a different nominal composition, and the reaction occurred at a different temperature. As suggested by Yamamoto [8] and by Ribeiro et al. [10], the addition of B$_4$C leads at the same reaction temperature to a higher solubility of pure Carbon than the addition of Carbon, which allows performing the reaction at lower temperatures: in the present work, a change in the lattice parameter is already observed at 720°C. This is the main reason for the enhanced $J_c$ values in the present work: the other authors added B$_4$C to MgB$_2$ bulk samples used considerably higher temperatures [9,10,17].

In our case, a deterioration of the transport properties of the conductors was observed at 900°C, which is attributed to reaction with the Fe matrix [13,14,15]. It is important to note that the used powders were mixed with an excess of B with respect to the stoichiometric ratio, thus probably compensating the B losses due to the interaction with the sheath. (at least up to 800°C) This fact is confirmed by the measurements on the B$_4$C doped wire with 0.657 mm diameter (Fig. 7). In this case, the critical current density $J_c$ was lower than that of the same conductor deformed to 1.11 mm, reacted at the same temperature (800°C). The influence of the reaction layer is expected to be more effective on the smaller wires.

It is known that MgB$_2$ wires with SiC nanopowder additions reach higher values of $J_c$, the highest being $10^4$ A/cm$^2$ at 12 T [18,19,20]. However it has to be noted that our B$_4$C powders has a considerably larger size (500 nm). In addition, no effort was undertaken to further densify the powder mixture in the filaments. The obtained results are promising for a further enhancement of the transport properties of the $B_4C$ added wires. It is expected that further improvements can be obtained by using a harder matrix e. g. Stainless Steel, by reducing the size of the filaments and by the optimization of the annealing conditions.

## Conclusions

In this study, we have investigated the transport properties of in situ $B_4C$ doped $MgB_2$ conductors. Measurements of transport properties revealed a marked enhancement of the current carrying capability of the wires after reaction at 800°C. The analysis of the reacted powder extracted by the wires revealed a shrinkage of the lattice parameter *a*, thus indicating a substitution of B by C. The present results confirm the potential role of $B_4C$, besides SiC, as an alternative addition for improving the transport properties of $MgB_2$ based conductors.

## Acknowledgements

This work is supported by the EU FP 6 project NMP-3-CT2004-505724.

**Figure captions**

**Table 1:** The lattice parameters *a* and *c* for different heat treatment conditions.

**Fig. 1:** Optical microscope image of the Fe sheathed wires of 1.11 mm diameter.

**Fig. 2:** X-ray diffraction patterns of the $MgB_2$ powder extracted from the filaments after reaction at various temperatures. The main $MgB_2$ peaks and the impurity phases are labelled.

**Fig. 3:** Optical microscope image of a cross section of a 1.11 mm diameter wire heat treated at 800°C, 1 hour, showing the reaction layer between the Fe sheath and the powder.

**Fig 4:** Lattice parameter *a* vs. reaction temperature showing a shrinking of the *ab* plane due to the interaction of the dopant $B_4C$ with the basic elements Mg and B.

**Fig. 5:** Critical current density $J_c$ vs. applied magnetic field *B* for the pure and the $B_4C$ doped conductor at various annealing temperatures.

**Fig. 6:** Critical current density $J_c$ vs. applied magnetic field *B* for the $B_4C$ doped wire after reaction at 800 and 900°C.

**Fig. 7:** Critical current density $J_c$ vs. applied magnetic field *B* at 4.2 and 20K for a $B_4C$ doped wire of 1.11 and 0.657 mm diameter, reacted 1 hour at 800°C.

**Fig. 8:** Comparison between the critical current density $J_c$ and the applied magnetic field *B* for various added $MgB_2$ wires and tapes with $B_4C$ and SiC addition.

**Table 1:** The lattice parameter $a$ and $c$ for different heat treatment conditions.

| $T_{reaction}$ (°C), time (min/ h) | $a$ (Å) | $c$ (Å) |
|---|---|---|
| 670°C, 2 min | 3.0854(3) | 3.5228(4) |
| 720°C, 20 min | 3.8016(4) | 3.5271(5) |
| 800°C, 60 min | 3.0797(5) | 3.5275(6) |

**Fig. 1:** Optical microscope image of the Fe sheathed wires of 1.11 mm diameter.

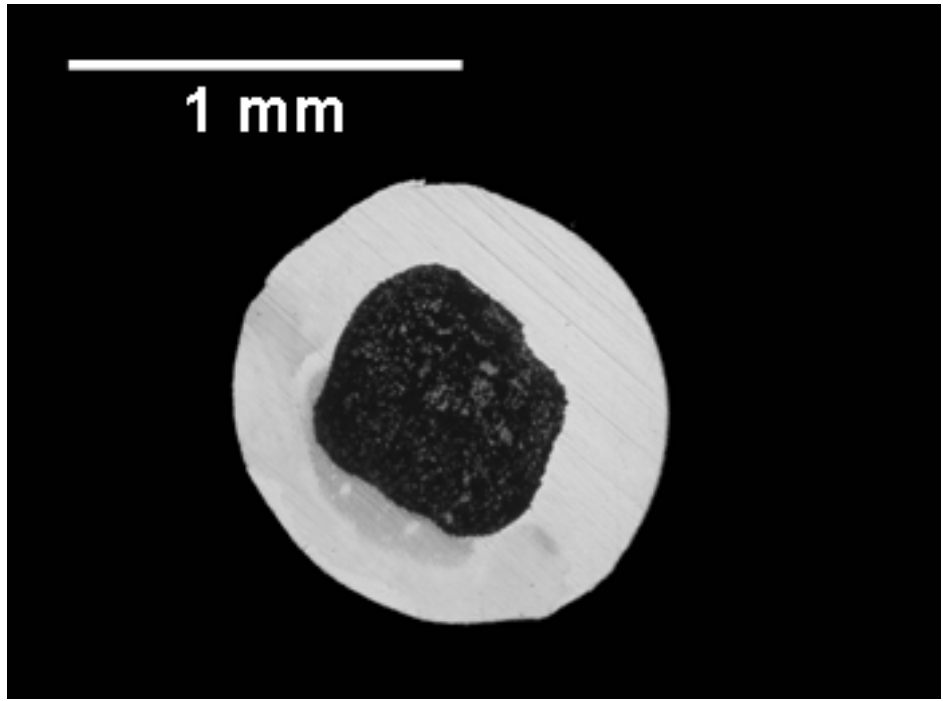

**Fig. 2:** X-ray diffraction patterns of the MgB$_2$ powder extracted from the filaments after reaction at various temperatures. The main MgB$_2$ peaks and the impurity phases are labelled.

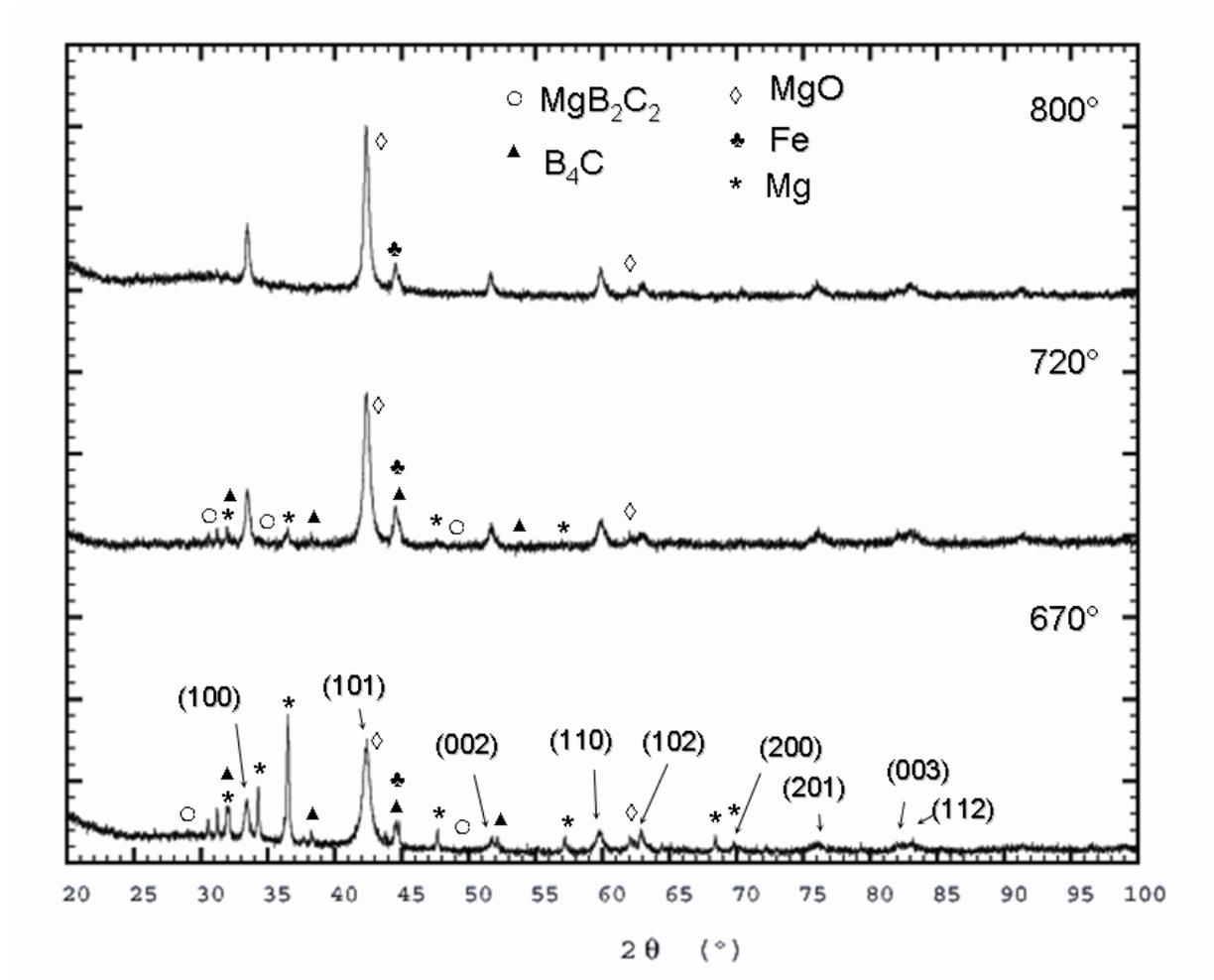

**Fig. 3:** Optical microscope image of a cross section of a 1.11 mm diameter wire heat treated at 800°C, 1 hour, showing the reaction layer between the Fe sheath and the powder.

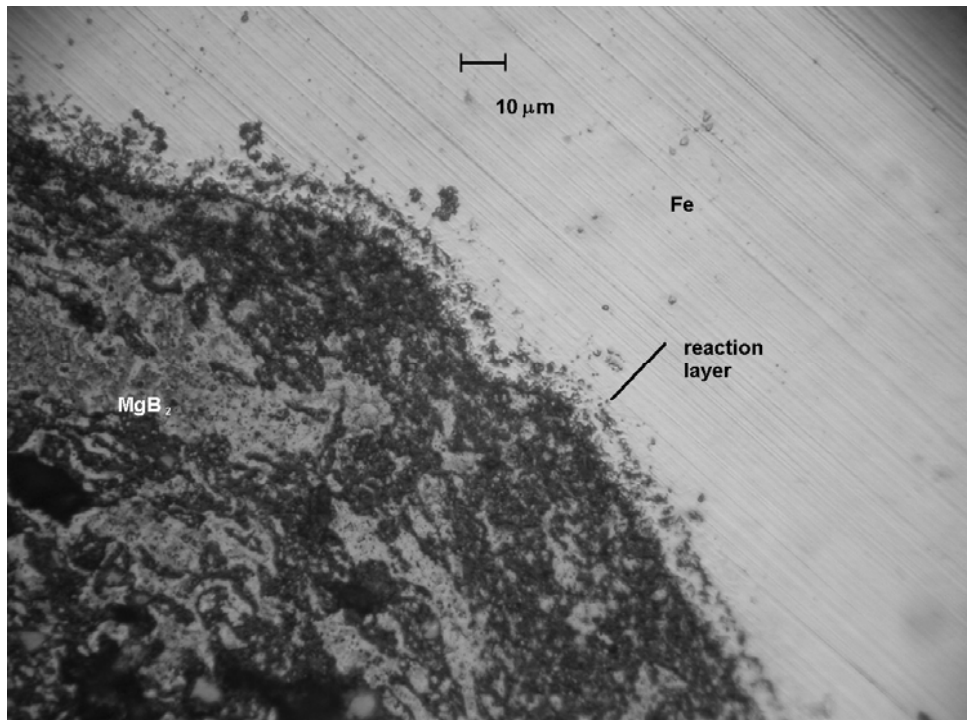

**Fig 4:** Lattice parameter *a* vs. reaction temperature showing a shrinking of the *ab* plane due to the interaction of the dopant B$_4$C with the basic elements Mg and B.

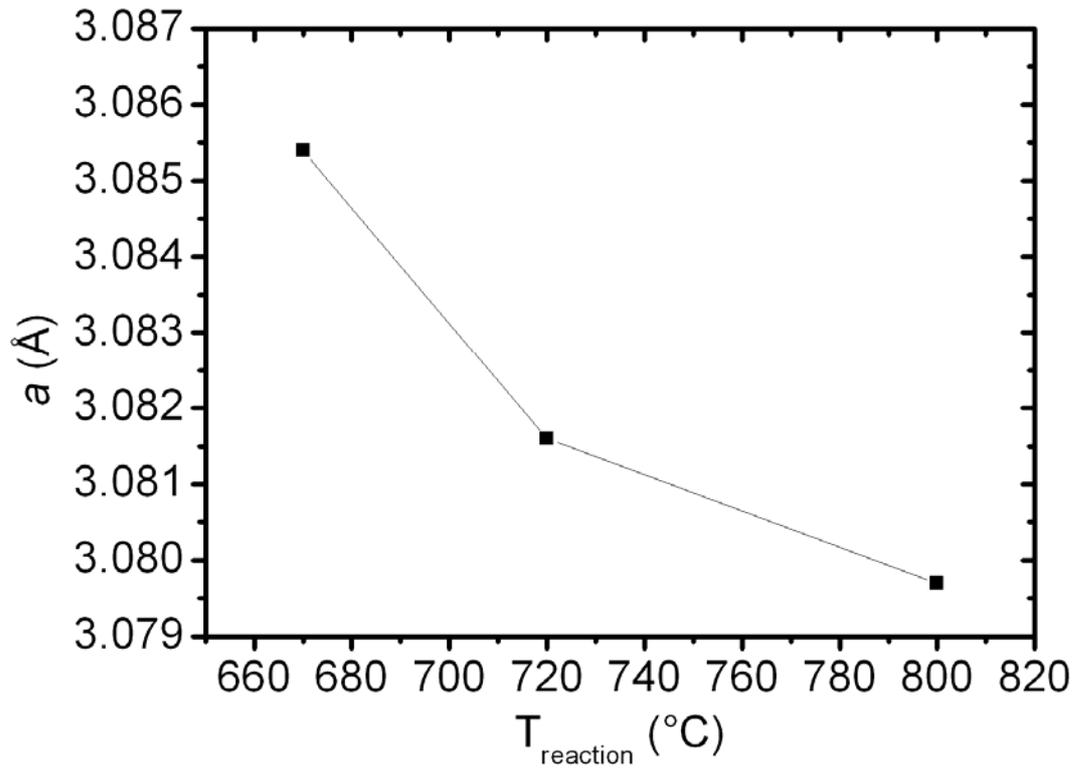

**Fig. 5:** Critical current density $J_c$ vs. applied magnetic field $B$ for the pure and the B$_4$C doped conductor at various annealing temperatures.

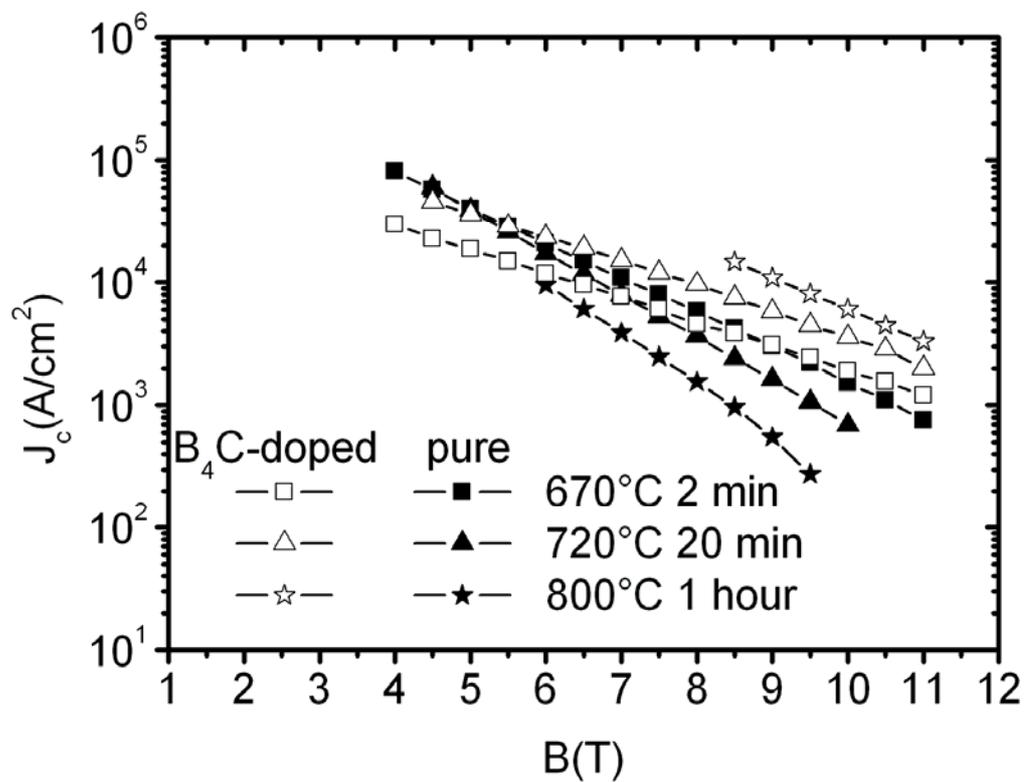

**Fig. 6:** Critical current density $J_c$ vs. applied magnetic field $B$ for the B$_4$C doped wire after reaction at 800 and 900°C.

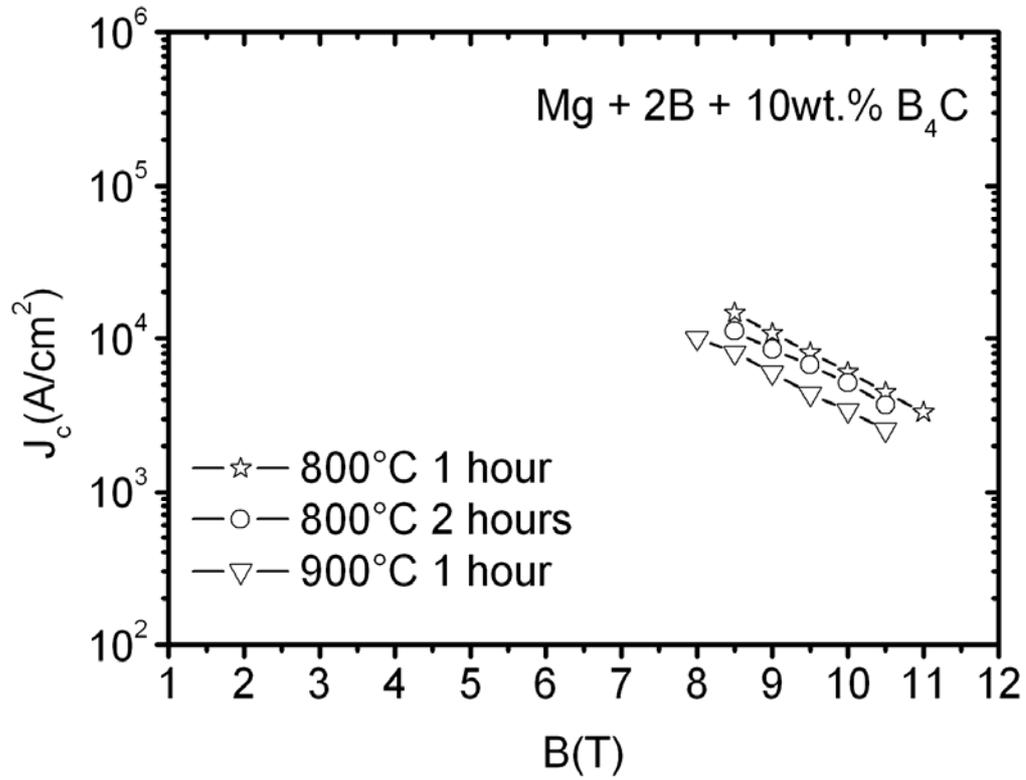

**Fig. 7:** Critical current density $J_c$ vs. applied magnetic field $B$ at 4.2 and 20K for a $B_4C$ doped wire of 1.11 and 0.657 mm diameter, reacted 1 hour at 800°C.

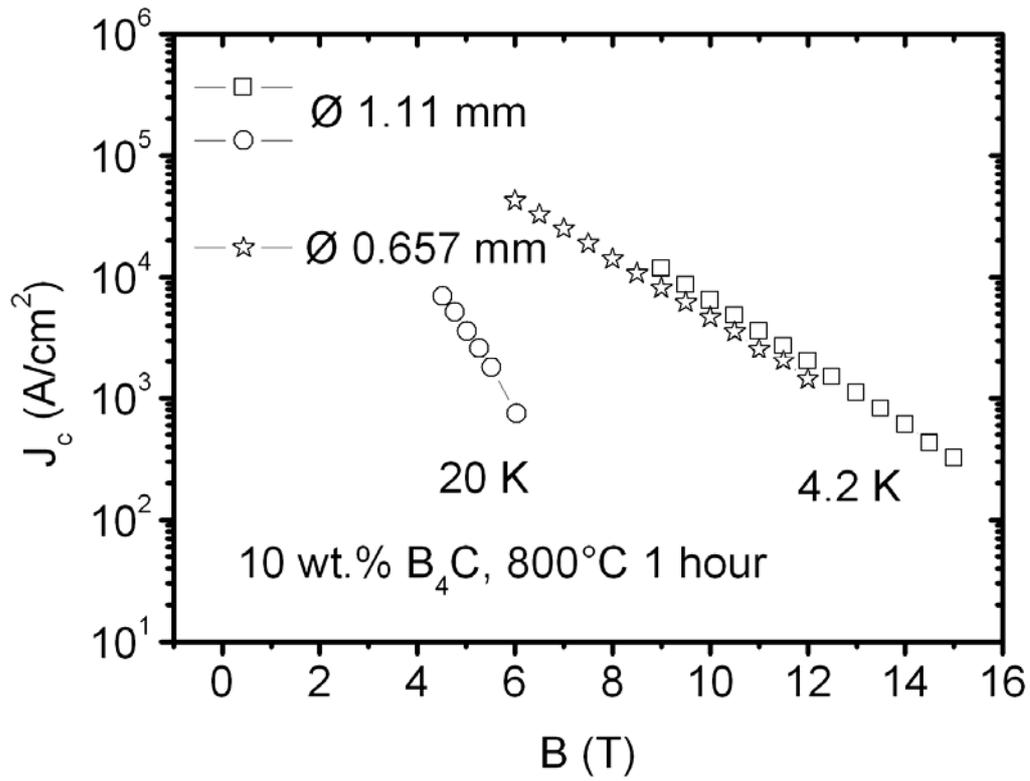

**Fig. 8:** Comparison between the critical current density $J_c$ and the applied magnetic field $B$ for various added MgB$_2$ wires and tapes with B$_4$C and SiC addition.

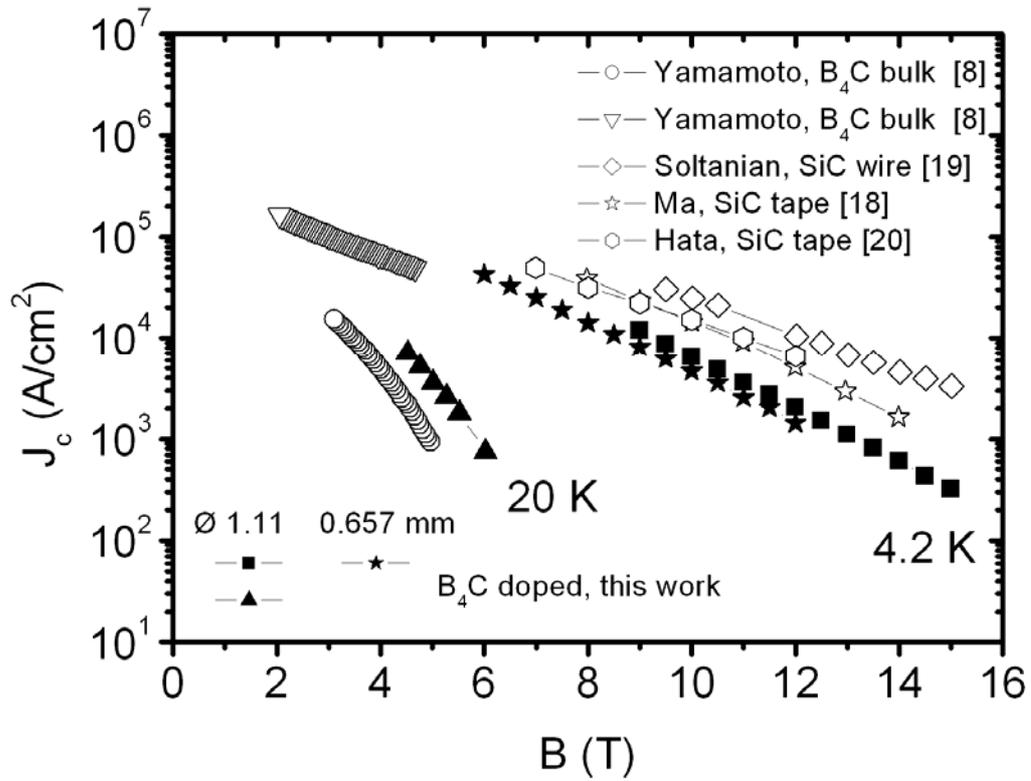